\RequirePackage{fixltx2e}
\documentclass[aip,jcp,reprint,showpacs]{revtex4-1}
\usepackage{amsmath,amsfonts} 
\usepackage{graphicx}
\usepackage[acronym]{glossaries} %
\usepackage{dcolumn} 
\usepackage{bm} 
\usepackage{ifpdf} 
\usepackage{natbib}
\usepackage{wasysym}
\usepackage[byname]{smartref}
\usepackage{color}
\usepackage{nicefrac}
\usepackage[breaklinks, %
colorlinks, %
linkcolor=blue, %
citecolor=blue, %
urlcolor=blue]{hyperref}
\usepackage[table]{xcolor}
\usepackage{multirow}
\usepackage{braket}
\usepackage{float}

\newcommand{\Tr}{\text{Tr}}
\newcommand{\Trn}{\text{Tr}_\text{n}}
\newcommand{\bff}[1]{\mathbf{#1}}
\newcommand{\bs}{\boldsymbol}
\newcommand{\fo}[1]{\frac{1}{#1}}
\newcommand{\be}{\begin{equation}}
\newcommand{\bea}{\begin{eqnarray}}
\newcommand{\ee}{\end{equation}}
\newcommand{\eea}{\end{eqnarray}}
\newcommand{\eq}[1]{Eq.~(\ref{#1})}
\newcommand{\nn}{\nonumber}

\graphicspath{{./pics/}}

\begin{document}

\newacronym{CI}{CI}{conical intersection} %
\newacronym{FC}{FC}{Franck-Condon} %
\newacronym{PES}{PES}{potential energy surface} %
\newacronym{DOF}{DOF}{degrees of freedom} %
\newacronym{LVC}{LVC}{linear vibronic coupling} %
\newacronym{QVC}{QVC}{quadratic vibronic coupling} %
\newacronym{BMA}{BMA}{2,6-bis(methylene)-adamantyl} %
\newacronym{MIA}{MIA}{2-methylene-6-isoproplidene-adamantyl} %
\newacronym{NFGR}{NFGR}{non-equilibrium Fermi golden rule} %
\newacronym{vMCG}{vMCG}{variational multiconfiguration Gaussian} %
\newacronym{MCTDH}{MCTDH}{multiconfiguration time dependant Hartree} %

\title{A perturbative formalism for electronic transitions through conical intersections in a fully quadratic vibronic model}
\author{Julia S. Endicott}
\affiliation{Department of Physical and Environmental Sciences,
  University of Toronto Scarborough, Toronto, Ontario, M1C 1A4,
  Canada}\affiliation{Chemical Physics Theory Group, Department of Chemistry, University of Toronto, Toronto,
  Ontario, M5S 3H6, Canada}
  
  \author{Lo{\"i}c Joubert-Doriol}
\affiliation{Department of Physical and Environmental Sciences,
  University of Toronto Scarborough, Toronto, Ontario, M1C 1A4,
  Canada}\affiliation{Chemical Physics Theory Group, Department of Chemistry, University of Toronto, Toronto,
  Ontario, M5S 3H6, Canada}

\author{Artur F. Izmaylov} %
\affiliation{Department of Physical and Environmental Sciences,
  University of Toronto Scarborough, Toronto, Ontario, M1C 1A4,
  Canada}\affiliation{Chemical Physics Theory Group, Department of Chemistry, University of Toronto, Toronto,  Ontario, M5S 3H6, Canada}
\date{\today}		

\begin{abstract}
We consider a fully quadratic vibronic model Hamiltonian for studying photoinduced electronic transitions through conical intersections. 
Using a second order perturbative approximation for diabatic couplings we derive an analytical expression for the time evolution of electronic populations at a given temperature. This formalism extends upon a previously developed perturbative technique for a linear vibronic coupling Hamiltonian. The advantage of the quadratic model Hamiltonian is that it allows one to use separate quadratic representations for potential energy surfaces of different electronic states and a more flexible 
representation of interstate couplings. 
We explore features introduced by the quadratic Hamiltonian in a series of 2D models, and then apply our formalism to 
the 2,6-bis(methylene) adamantyl cation, and its dimethyl derivative. The Hamiltonian parameters for the molecular systems have been obtained from electronic structure calculations followed by a diabatization procedure. 
The evolution of electronic populations in the molecular systems using 
the perturbative formalism shows a good agreement with that from variational quantum dynamics.  
\end{abstract}
\maketitle

\section{Introduction}
\sloppy
Photoinduced charge, proton, and energy transfers  are quite common processes in many areas of biological~\cite{Zhong:2000/PNAS/14056,Chattoraj:1996/PNAS/8362, Miskovsky:2002/IJP/45} and technological~\cite{Book/Haddon:1987,Parthenopoulos:1991/JPC/2668,Moller:1998/CPL/291} significance. An adequate description of these processes requires inclusion of multiple electronic states and thus goes beyond the Born-Oppenheimer approximation. \Glspl{PES} of electronic states in large systems quite commonly intersect forming \glspl{CI}~\cite{Yarkony:1996/rmp/985,Migani:2004/271} which provide an efficient channel for energy redistribution via non-adiabatic dynamics.~\cite{Hahn:2000/jpcb/1146,polli:2010/nature/440} Generally, to account for quantum effects associated with non-adiabatic dynamics requires using methods of quantum dynamics whose computational cost scales exponentially with the number of \gls{DOF}.~\cite{Meyer2009:mctdh} 
Mixed quantum-classical approaches\cite{Tully:1990/jcp/1061,Subotnik:2011bo,Bittner:1995vx,Kapral:1999/jcp/8919} 
partially alleviate the problem by treating nuclear dynamics with classical mechanics 
and employing quantum consideration to accommodate non-adiabatic transitions. However, application of these approaches 
can also be computationally demanding 
if one accounts for increasing number of classical trajectories needed for adequate sampling 
of quantum transitions and computational cost of electronic \gls{PES} calculations associated with each trajectory.

For practical purposes though, one does not need to know dynamics of all \gls{DOF} in a large system undergoing charge or energy transfer. In many cases the main interest is in properties related to only the electronic \gls{DOF}: rates of electronic transitions or electronic state population dynamics.  
However, one cannot simply disregard nuclear \gls{DOF}  because their dynamics is the main cause of electronic transitions 
in the non-adiabatic dynamics. Instead, at least in semi-rigid systems, one can parametrize the nuclear \gls{DOF} with a simple exactly solvable harmonic model and couple its nuclear dynamics in an analytic form with electronic \gls{DOF} through electron-nuclear couplings treated perturbatively in the spirit of the Marcus and F\"orster 
theories.~\cite{Book/Nitzan:2006,Book/Zwanzig:2001} Such perturbative treatment results in effective time-dependent electron transition rates that originate from the nuclear motion and define the electronic dynamics. 
Using the exactly solvable model for the nuclear dynamics makes these transition rates amenable to analytical treatment and thus removes the burden of the numerical propagation of the nuclear \gls{DOF} completely. 

These ideas have been implemented recently in the so-called \gls{NFGR}\cite{Izmaylov:2011/jcp/234106} and generalized 
master equation approaches\cite{Pereverzev:2006/JCP/104906} for \glspl{CI} of two electronic states parametrized within the \gls{LVC} model Hamiltonian~\cite{Domcke2004:conical,Koppel:1984/ACP/59,Worth:2004/ARPC/127}
\bea \notag
H_{\rm LVC}&=&\sum_{i=1}^{N} \left( \begin{array}{cc}
(p_{i}^{2}+\Omega_{i}^{2}q_{i}^{2})/2& 0\\
0 & (p_i^{2}+\Omega_{i}^{2}q_{i}^{2})/2 +\Delta E
\end{array} \right)\\ \label{eq:LVC}
&+&\left( \begin{array}{cc}
d_{D,i}q_{i} & c_{i}q_{i}\\
c_{i}q_{i} & d_{A,i}q_{i}
\end{array} \right).
\eea
$H_{\rm LVC}$ represents coupled donor and acceptor diabatic states by $N$-dimensional harmonic oscillators with frequencies $\Omega_i$, mass weighted coordinates $q_i$ and corresponding momenta $p_i$. Although the surfaces have the same frequencies they have different linear shifts, $d_{D,i} \ne d_{A,i}$, and are energetically separated by $\Delta E$. 
The surfaces are coupled by linear coupling terms $c_i q_i$ that give rise to the \gls{CI} topology 
in the adiabatic representation. 

\begin{figure}
\includegraphics[width=0.4\textwidth]{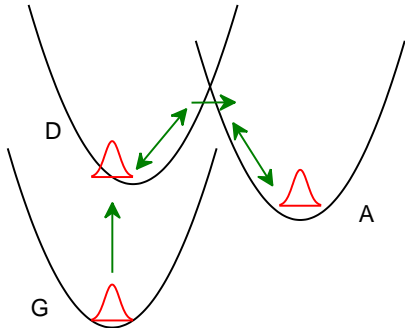}
\caption{Photo-induced non-radiative transition through a conical intersection.}
\label{fig:system}
\end{figure}

Assuming only constant shift differences $x_{G,i}$ between minima of the donor state and some ground electronic state 
with the Hamiltonian
\bea	\label{eq:HG}
\tilde{H}_G = \sum_{i=1}^N \frac{1}{2}\left[ p_{i}^{2}+\Omega_{i}^{2}(q_{i}+x_{G,i})^{2}\right]
\eea
using the \gls{NFGR} approach one can study non-equilibrium electronic population dynamics when an initial ultra-fast laser pulse excites 
a ground state Boltzmann density to the donor state and then the electronic population is transferred to the acceptor state via non-radiative transition through a \gls{CI} (see Fig.~\ref{fig:system}).
As a limiting case of the zero shift between the ground and donor states the \gls{NFGR} formalism can also be applied to model radiationless transfer starting with the equilibrium Boltzmann density on the donor state. 
Using the \gls{NFGR} approach the electronic population dynamics of excited states were calculated for some test molecules and showed quantitative agreement for short times and qualitative agreement for longer times when compared to results 
of variational wave packet techniques.~\cite{Izmaylov:2011/jcp/234106}
\sloppy

The assumption of the same frequencies and normal modes for the ground, donor, and acceptor states limits the 
applicability of the \gls{NFGR} method. 
An extension of the \gls{LVC} Hamiltonian, which is more general and allows the inclusion of effects such as the 
Duschinsky rotation~\cite{Duschinsky:1937/apussr/132}  of the normal coordinates and changes in vibrational frequencies between two states, 
is the \gls{QVC} model Hamiltonian
\bea\label{eq:QVC}
H_{\rm QVC}&=&\left( \begin{array}{cc}
H_{D}& V_{DA}\\
V_{AD}& H_{A}
\end{array} \right),\\
H_{D}&=&\sum_{i=1}^{N}\frac{1}{2}p_{D,i}^{2}+\frac{1}{2}\Omega_{D,i}^{2}q_{D,i}^{2},\\
H_{A}&=&\sum_{i=1}^{N}\frac{1}{2}p_{A,i}^{2}+\frac{1}{2}\Omega_{A,i}^{2}q_{A,i}^{2} +\Delta E,\\ \label{eq:QVCe}
V_{DA}&=&V_{AD}=\sum_{i,j=1}^{N} \Theta_{ij}q_{D,i}q_{D,j}\\ \notag
&&+\sum_{i=1}^{N}\gamma_{i}q_{D,i}+\Delta_{DA},
\eea
where harmonic frequencies of the donor $\Omega_{D,i}$ and acceptor $\Omega_{A,i}$ states can be different, and  
normal modes of the acceptor state $q_{A,i}$ can be written as shifted and rotated normal modes of the donor state $q_{D,i}$
\bea
\label{eq:AcceptRot}
q_{A,i} &=& \sum_{j=1}^{N}J_{ij}^{(A)}(q_{D,j}+x_{A,j}).
\eea
Here we introduced the unitary Duschinsky matrix $J_{ij}^{(A)}$ and the shift vector $x_{A,j}$ 
between the minima of the acceptor and donor states. 
Using the Duschinsky matrix, a Hessian matrix of the acceptor state can be written in terms of the donor 
normal modes as $\bs\Omega_{A}' = \bff{J^{(A)}}^{\dagger}\bs\Omega_{A}\bff{J^{(A)}}$. 
$H_{\rm QVC}$  also differs from $H_{\rm LVC}$ by quadratic  ($\Theta_{ij}q_{D,i}q_{D,j}$), 
and constant ($\Delta_{DA}$) terms in the coupling potentials $V_{DA}$ and $V_{AD}$. 
Picconi {\it et al.}~\cite{Picconi:2012/jcp/244104} showed that using the \gls{QVC} model rather than the \gls{LVC} model can cause significant changes to both calculated spectra and electronic population dynamics in thymine. 
The \gls{QVC} Hamiltonian with only a constant interstate coupling $V_{DA} = V_{AD} = \Delta_{DA}$ 
have been considered by Borrelli {\it et al.},\cite{Borrelli:2011/PCCP/4420} however, 
this set up cannot result in the CI topology.  
As has been repeatedly shown, ignoring topological features of the \gls{CI} 
can lead to qualitatively different predictions for nuclear dynamics.\cite{Ryabinkin:2013/prl/220406,Joubert:2013/jcp/234103,Schon:1994/cpl/55}

Building the \gls{QVC} Hamiltonian from {\it ab initio} electronic structure
calculations is not a straightforward task because these calculations provide 
the adiabatic picture while the \gls{QVC} Hamiltonian is written in the diabatic representation. 
The exact diabatic representation is impossible in practice for systems with more than 
one \gls{DOF},\cite{Mead:1982/JCP/6090} and therefore, 
here we use approximate, so-called regularized diabatic states proposed 
by K\"{o}ppel and coworkers.\cite{Koppel:2001/JCP/2377,Koppel:2010/MP/1069} 
Construction of these states is based on removing only leading terms of non-adiabatic couplings that cause 
singularities at the \gls{CI} seam. This removable part involves only states which participate in the \gls{CI}, 
thus it is possible to rotate adiabatic states into regularized diabatic states where
the singularities disappear. 

In this paper we focus on extending the \gls{NFGR} approach to the \gls{QVC} Hamiltonian.   
To illustrate efficiency and accuracy of the developed 
approach we investigate internal electron transfer in the \gls{BMA} and \gls{MIA} cations and compare 
our results to those obtained using the \gls{vMCG}\cite{Worth:2004/fd/307} and \gls{MCTDH}\cite{mctdh} methods. 
 
The rest of the paper is organized as follows. Section \ref{sec:meth} provides main steps in the derivation of the central equation used to calculate evolution of the electronic donor state population, while further details are given in the Appendix.  Section \ref{sec:results} uses 2D models and parametrized \gls{BMA} and \gls{MIA} model Hamiltonians to explore the capabilities 
of the \gls{NFGR} method. Finally, Section \ref{sec:con} concludes and provides an outlook for future work. Atomic units are used 
throughout this paper.

\section{Method}
\label{sec:meth}
For the \gls{QVC} Hamiltonian [\eq{eq:QVC}] the electronic population of the donor diabatic state $\ket{D}$ can be written 
as a projection of the full electron-nuclear density on the donor state traced over all nuclear \gls{DOF},
\begin{equation}\label{eq:start}
P_D(t)=\Tr_{n}\left[\left<D\left|e^{-iH_{\rm QVC}t}\rho(0)e^{iH_{\rm QVC}t}\right|D\right>\right].
\end{equation} 
To treat the quantum propagator $e^{-iH_{\rm QVC}t}$ we employ a perturbation theory expansion 
by splitting the $H_{\rm QVC}$ Hamiltonian as
\bea\label{eq:Ham}
H_{\rm QVC} &=& H_0 + V,
\eea
with
\bea
H_{0} &=& \left|D\left>\right<D\right|H_{D}+ \left|A\left>\right<A\right|H_{A},\\
V &=&\left|D\left>\right<A\right|V_{DA}+\left|A\left>\right<D\right|V_{AD},
\eea
where $H_D$, $H_A$, $V_{DA}$, and $V_{AD}$ are given by Eqs.~(\ref{eq:QVC}-\ref{eq:QVCe}).
The perturbative expansion of the propagator in the interaction picture up to the second order in $V$
gives the evolution operator 
\begin{equation}
U(t) = 1-\int_{0}^{t} \mathrm{d}t'\int_{0}^{t'} \mathrm{d}t'' e^{iH_{0}t'}Ve^{-iH_{0}(t'-t'')}Ve^{-iH_{0}t''}.\\
\end {equation}
Assuming an ultrafast laser pulse that promotes the ground state density $\ket{G}\rho_n\bra{G}$ to the initial density 
on the donor state $\rho(0) = \ket{D}\rho_n(0)\bra{D}$ and using 
the perturbative evolution operator $U(t)$ for the propagator, the equation for the donor state population becomes
\begin{equation}\label{eq:PT}
\begin{aligned}
\tilde P^{(2)}_{D}(t)=&\Trn\left[\left<D\left|U(t)\right|D\right>\rho_{n}(0)\left<D\left|U^{\dagger}(t)\right|D\right>\right]. \\
\end{aligned}
\end {equation}
We further simplify Eq.~(\ref{eq:PT}) by excluding all terms of higher than the 2$^{\rm nd}$ order in $V$
and introducing $\rho_{n}(0)=e^{-\beta H_G}/{\rm Tr}[e^{-\beta H_G}]$ as a Boltzmann distribution 
of the ground state with inverse temperature $\beta=1/(k_{B}T)$ 
\bea\label{eq:popsimp}
P_D^{(2)}(t)&=&1-2\mathrm{Re}\int_{0}^{t} \mathrm{d}t'\int_{0}^{t'} \mathrm{d}t'' f(t',t''),
\eea
where the time-correlation function $f(t',t'')$ is 
\bea\notag
f(t',t'')&=&\frac{1}{\Trn[e^{-\beta H_{G}}]}\Trn\Big[ e^{-\beta H_{G}}e^{iH_{D}t'}\\\label{eq:fTr}
&&\times V_{DA}e^{-iH_{A}(t'-t'')} V_{AD}e^{-iH_{D}t''}\Big] .
\eea
Here, the ground state Hamiltonian $H_G$ is defined in a more general form than in the \gls{LVC} consideration [cf. \eq{eq:HG}] 
\bea
H_{G}&=&\sum_{i=1}^{N}\frac{1}{2}p_{G,i}^{2}+\frac{1}{2}\Omega_{G,i}^{2}q_{G,i}^{2},\\
q_{G,i}&=& \sum_{j=1}^{N}J_{ij}^{(G)}(q_{D,j}+x_{G,j}),
\eea
where $q_{G,i}$ are corresponding normal modes that are generally rotated by $J_{ij}^{(G)}$ and shifted by $x_{G,j}$ 
with respect to the donor normal modes.

Analytical form of the $f(t',t'')$ function is obtained using Gaussian integration for the traces in Eq.~(\ref{eq:fTr}), 
details of the derivation can be found in the Appendix.
The $f(t',t'')$ function contains all parameters of the $H_{\rm QVC}$ and $H_G$ Hamiltonians 
as well as temperature of the initial Boltzmann distribution. Although the explicit form of the
$f(t',t'')$ function is very lengthy (see the Appendix), this function encompasses a few essential factors 
for the electronic population dynamics: (1) couplings between vibrational levels of 
the donor and acceptor states, (2) zeroth order energy differences between interacting vibrational levels, 
(3) initial conditions of the nuclear distribution. A significant role in modulating the couplings between 
the vibrational levels is played by the corresponding \gls{FC} overlaps between the associated 
nuclear wave-functions. Due to an exponential dependence on the distance between the donor-acceptor 
minima the \gls{FC} overlap influence usually overpowers that of the polynomial terms from the $V_{DA}$ potential.   

If vibrational levels of uncoupled donor and acceptor states become close in energy 
the population calculated using Eq.~(\ref{eq:popsimp}) becomes negative at longer times. 
To avoid this unphysical behaviour we perform a partial infinite resummation of the perturbative population 
expansion using the corresponding cumulant expansion.\cite{Izmaylov:2011/jcp/234106} 
In the second order, the cumulant expansion amounts to exponentiating the 
second term in Eq.~(\ref{eq:popsimp})
\begin{equation}
\label{eq:cumulant}
P_D^{[2]}(t)=e^{-2\text{Re}\left[ \int_{0}^{t} \mathrm{d}t'\int_{0}^{t'} \mathrm{d}t'' f(t',t'') \right]}.
\end {equation}
Using the analytical expression for the $f(t',t'')$ function evaluating the electronic population dynamics
in Eq.~(\ref{eq:cumulant}) requires only two-dimensional time integration that is done numerically. 
As a result the \gls{NFGR} procedure scales quadratically with the number of time steps and 
cubically with the number of nuclear \gls{DOF} due to matrix manipulations involved in the $f(t',t'')$ evaluation. 

\section{Results and Discussions}
\label{sec:results}

\subsection{2D Model}
\label{sec:2D}

\begin{figure}
\centering
\includegraphics[width=0.5\textwidth]{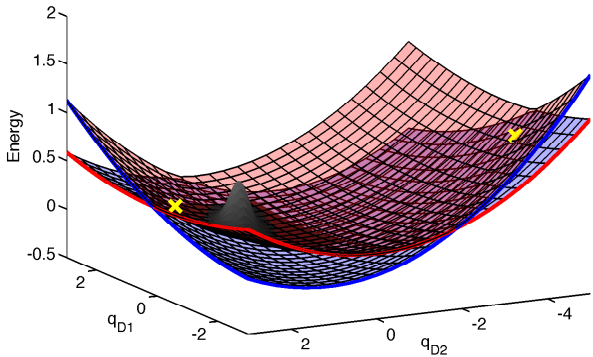}
\caption{Adiabatic \protect\glspl{PES} for the generic 2D model given in Table \ref{tab:param}. 
The lines at the edges of the plot show the diabatic states: the donor state in red and the acceptor state in blue.
The crosses show the two conical intersections at (-1.3,-4.4) and (0.6,2.2). 
The wave packet is a scaled version of the initial population density on the donor surface.
}
\label{fig:2D}
\end{figure}

First, we will illustrate properties of the \gls{NFGR} approach applied to 2D $H_{\rm QVC}$ model Hamiltonians with various parameters. The parameters for our generic 2D model are given in Table \ref{tab:param}, 
we will use this system as our base and then explore some modifications of its parameters. 
The system consists of ground, donor, and acceptor states, which can have different minima, 
frequencies, and normal modes. Figure  \ref{fig:2D} shows \glspl{PES} of the donor and acceptor states and the initial population density after excitation from the ground state. Our results will be compared to those of the exact quantum dynamics obtained using the split operator method.\cite{Tannor2007:intro} 
In all \gls{NFGR} simulations temperature is $2\times10^{-8}$ a.u. (5$\times 10^{-3}$K), 
this makes values of $\beta$ very large but finite and allows us to compare \gls{NFGR} simulations 
with those using the split operator approach at zero temperature. 

\begin{table*}\centering
\caption{Parameters for the generic 2D model.}
\begin{tabular}{cccccccc}
\firsthline
\hline
$\bs\Omega_{D}$&$\bs\Omega'_{G},\bs\Omega'_{A}$ &$\bff{x}_{G}$&$\bff{x}_{A}$& $\bs\Theta$ & $\bs\gamma$ & $\Delta_{DA}$ &$\Delta E$\\
\hline
$\left(\begin{array}{cc}
0.2& 0.0\\
0.0& 0.3
\end{array} \right)$&
 $\left(\begin{array}{cc}
0.2& 0.1\\
0.1& 0.3
\end{array} \right)$ & 
$\left(\begin{array}{c}
-0.4\\
-0.8
\end{array} \right)$ &
$\left(\begin{array}{c}
0.2\\
0.3
\end{array}\right) $ &
$\left(\begin{array}{cc}
0.0002& 0.0001\\
0.0001& 0.0003
\end{array} \right)$ & 
$\left(\begin{array}{c}
0.003\\
0.001
\end{array}\right) $ &
0.001&
-0.2\\
\hline\hline
\end{tabular}
\label{tab:param}
\end{table*}

\begin{figure}
\centering
\includegraphics[width=0.5\textwidth]{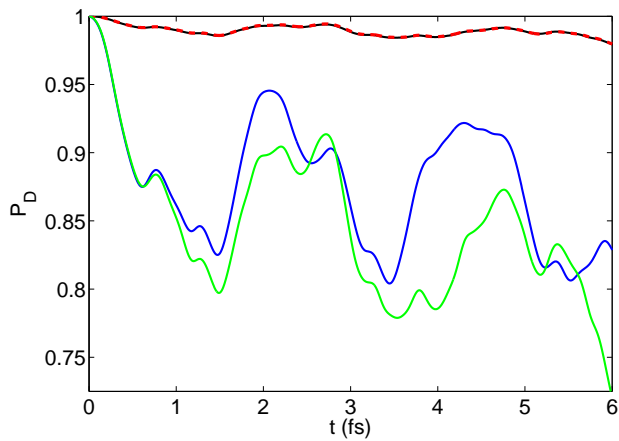}
\caption{Population dynamics using the exact and NFGR approaches for the generic 2D model (Table \ref{tab:param}): 
exact (solid black) and \protect\gls{NFGR} (dashed red); 
and for the strong coupling case (all couplings in Table \ref{tab:param} are enhanced 4 times): 
exact (solid blue), \protect\gls{NFGR} (green).}
\label{fig:coup}
\end{figure}

In Fig.~\ref{fig:coup} the population of the donor state over time for the system given in 
Table \ref{tab:param} is shown for two cases: 1) the weak coupling case, 
 where $\bs\Theta$, $\bs\gamma$ and $\Delta_{DA}$ are from Table \ref{tab:param}; and 2) the strong coupling case, where electronic coupling parameters $\bs\Theta$, $\bs\gamma$ and $\Delta_{DA}$ are four times those given in Table \ref{tab:param}. 
The weak coupling case shows excellent agreement between results of the exact and \gls{NFGR} methods. 
As expected for the perturbative approximation, in the strong coupling case, this agreement is quantitative 
only for short times but it remains qualitative for later times. 

\begin{figure}
\includegraphics[width=0.5\textwidth]{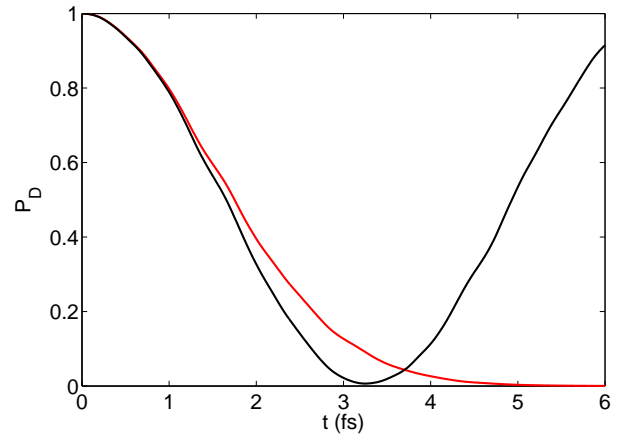}
\caption{Population dynamics for the generic 2D model (Table \ref{tab:param}) with 
modified parameters to satisfy the resonance condition ${\bs\Omega}'_{A}=\left(\begin{smallmatrix}
0.2& 0\\
0& 0.3
\end{smallmatrix} \right)$, $\Delta E=0$, and $\Delta_{DA}=0.01$: 
exact (black) and \gls{NFGR} (red).}
\label{fig:res}
\end{figure}

The time-scale and extent of the donor-acceptor population transfer is highly dependant on the relative 
energies of the vibrational levels in each diabatic state.
We will consider two cases: 1) when uncoupled vibrational levels of both diabates are 
energetically aligned, the resonance case; 
2) when there is an energy difference between these levels, the non-resonant case.
In the resonant case, the population of the donor state oscillates between 0 and 1 with a frequency 
that depends on the coupling strength. Fig.~\ref{fig:res} illustrates that although the \gls{NFGR} 
approach does not capture coherent population oscillations it does reproduce accurately the time-scale 
of the forward donor-acceptor transfer. \gls{NFGR} does not account for the back transfer, and therefore 
it plateaus after the donor population depletion. Using a master equation framework~\cite{Book/Breuer:2002} 
with \gls{NFGR} time-correlation functions [e.g., Eq.~(\ref{eq:Kubo})] can partially alleviate this drawback.

\begin{figure}
\centering
\includegraphics[width=0.5\textwidth]{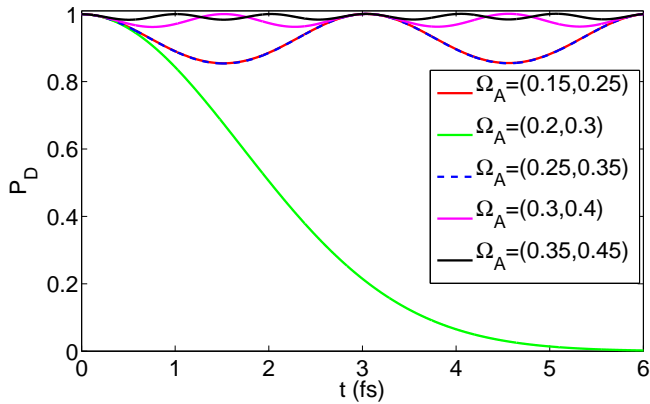}
\caption{\protect\gls{NFGR} population dynamics for the 2D model 
with $\Delta E$, $\bs\Theta$, $\bs\gamma$, $\bff{x_G}$, and $\bff{x_A}$ are 
all set to zero and $\bff{\Omega_{G}}=\bff{\Omega_{D}}=\left(\begin{smallmatrix}
0.2& 0\\
0& 0.3
\end{smallmatrix} \right)$, $\Delta_{DA}$=0.01 and $\bs\Omega_{A}$ is 
diagonal and varied according to the legend.}
\label{fig:Hasq} 
\end{figure}

In the non-resonant case,  instead of the complete population 
transfer we observe only small amplitude population oscillations 
(see Fig.~\ref{fig:coup}). With the \gls{LVC} Hamiltonian model used in previous 
work~\cite{Izmaylov:2011/jcp/234106}, the non-resonant regime could only appear due to the $\Delta E$ term.
With the \gls{QVC} model, the donor and acceptor states can have different frequencies and therefore be 
non-resonant for any $\Delta E$.  Figure~\ref{fig:Hasq} illustrates population oscillations arising 
from differences in donor-acceptor frequencies. 

\begin{figure}
\includegraphics[width=0.5\textwidth]{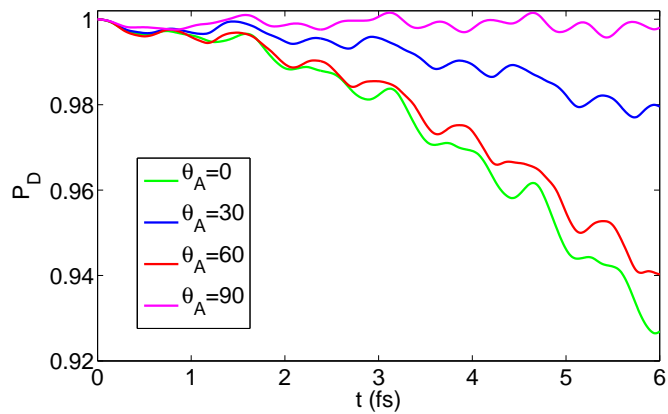}
\caption{\protect\gls{NFGR} population dynamics for 2D models 
with different values of the acceptor state rotation angle $\theta_A$ [\eq{eq:deftheta}]
and other parameters from Table \ref{tab:param}}.
\label{fig:Ha_rot} 
\end{figure}
\begin{figure}
\includegraphics[width=0.5\textwidth]{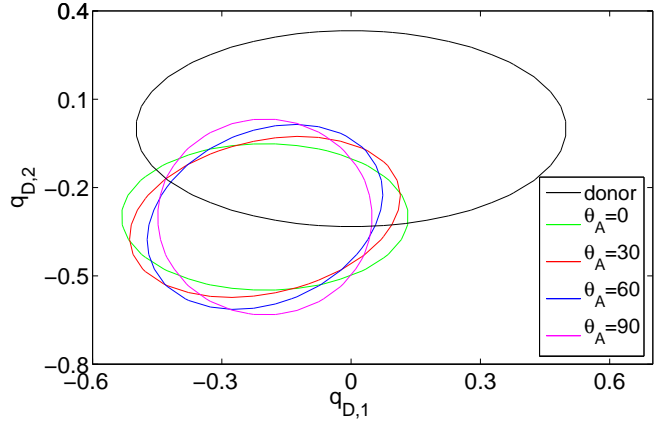}
\caption{Isoenergetic cross sections of the diabatic donor and acceptor \glspl{PES} for 2D models
with various values of the acceptor state rotation angle $\theta_A$  [\eq{eq:deftheta}] and other parameters 
from Table \ref{tab:param}.}
\label{fig:Ha_el} 
\end{figure}

Using the \gls{QVC} Hamiltonian, the \gls{NFGR} approach can also account for 
the rotation and translation of the vibrational normal modes of different electronic states 
(the Duschinsky effect). To study the effect of the Duschinsky rotation, we parametrized ${\mathbf J}^{(A)}$ and 
${\mathbf J}^{(G)}$ matrices as
\bea
\label{eq:deftheta}
{\mathbf J}^{(S)}=\left(\begin{matrix}\cos\theta_{S}&-\sin\theta_{S}\\\sin\theta_{S}&\cos\theta_{S}\end{matrix}\right),
\eea
where $S$ indicates the electronic state, and $\theta_{A}(\theta_{G})$ is the angle between 
normal modes of the donor and acceptor (ground) state.
Figure \ref{fig:Ha_rot} shows that the population transfer decreases with $\theta_A$. This can be attributed to the reduction 
of \gls{FC} overlaps between vibrational states of the donor and acceptor states with $\theta_A$ increase (see Fig.~\ref{fig:Ha_el}). 

\begin{figure}
\includegraphics[width=0.5\textwidth]{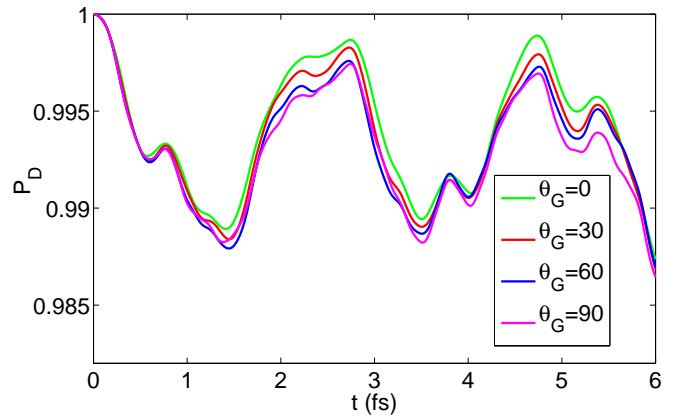}
\caption{\protect\gls{NFGR} population dynamics for 2D models 
with various values of the ground state rotation angle $\theta_G$ [\eq{eq:deftheta}] 
and other parameters from Table \ref{tab:param}.}
\label{fig:Hg}
\end{figure}
\begin{figure}
\includegraphics[width=0.5\textwidth]{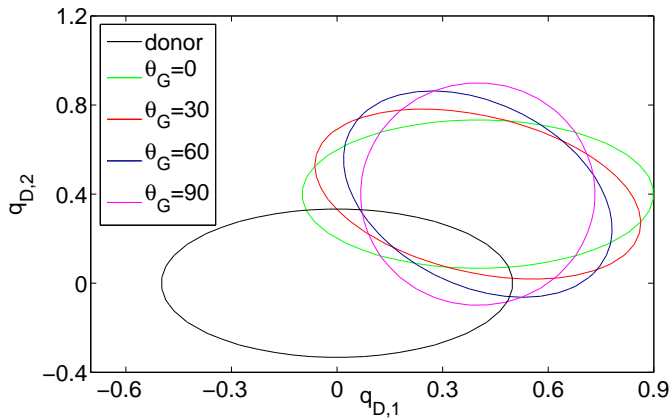}
\caption{Isoenergetic cross sections of the diabatic ground and donor \glspl{PES} for 2D models
with various values of the ground state rotation angle $\theta_G$ [\eq{eq:deftheta}] and other parameters 
from Table \ref{tab:param}.}
\label{fig:Hg_el} 
\end{figure}

The Duschinsky rotation $\mathbf{J}^{(G)}$ also affects the population dynamics because the initial nuclear 
distribution is taken as a Boltzmann distribution of the ground state in \eq{eq:PT} (see Figs. \ref{fig:Hg} and~\ref{fig:Hg_el}). 
The change in population transfer due to $\mathbf{J}^{(G)}$ is much smaller than that from $\mathbf{J}^{(A)}$ because the 
former does not affect coupling between donor and acceptor states but only modifies initial conditions.  
 Although the changes in population transfer with rotation of the ground state are small there is an overall trend of increased 
 population transfer with ground state rotation in this system.

\begin{figure}
\includegraphics[width=0.5\textwidth]{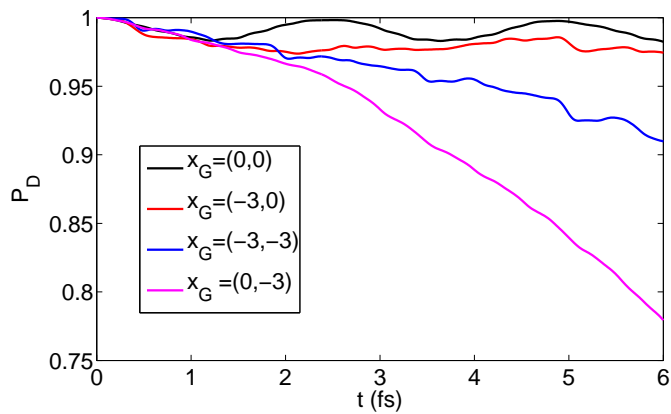}
\caption{\protect\gls{NFGR} population dynamics for 2D models 
with various values of the ground state shift $x_{G}$ and other parameters from Table \ref{tab:param}.}
\label{fig:gs}
\end{figure}

Besides rotations, the $H_{\rm QVC}$ Hamiltonian accounts for shifts between minima of different electronic states. 
Shifting the acceptor state minimum away from that of the donor state reduces the transfer as 
\gls{FC} overlaps between donor and acceptor vibrational states decrease. 
Increasing the distance between the ground and donor state minima puts a non-equilibrium initial population distribution
higher on a slope of the donor diabatic state after vertical excitation from the minimum of the ground state. 
Figure \ref{fig:gs} shows the electronic population dynamics for a series of systems with different ground state shifts. 
The generic model given in Table \ref{tab:param} has been modified for this part by making 
 $q_{D,1}$ the coupling mode with $x_{A,1}=0$, and $q_{D,2}$ the tuning mode with 
 $\gamma_{2}=\Theta_{2,2}=\Theta_{1,2}=\Theta_{2,1}=0$.
 This modification allows us to separate the effect from shifting the  
ground state minimum along the tuning mode from that along the coupling mode. 
Shifting the ground state minimum along the coupling coordinate increases slightly the population transfer 
[$\bff{x_G}=(-3,0)$ in Fig.~\ref{fig:gs}]. It is expected because as the wave packet travels further 
along the coupling coordinate it spends more time in areas of higher coupling. 
Shifting the ground state minimum along the tuning coordinate increases drastically the population 
transfer [$\bff{x_G}=(0,-3)$ in Fig.~\ref{fig:gs}].
This is consistent with the topography of the surfaces in Fig.~\ref{fig:2D},
 since in this case the wave packet oscillates between the two conical intersections (see Fig.~\ref{fig:2D}).
Shifting in both tuning and coupling directions [$\bff{x_G}=(-3,-3)$ in Fig.~\ref{fig:gs}] 
gives rise to a stepwise progression of the population transfer, which is a result of the wave 
packet oscillating between areas of low and high couplings. 

\subsection{\protect\gls{BMA} and \protect\gls{MIA} cations}

\begin{figure}
\includegraphics[width=0.5\textwidth]{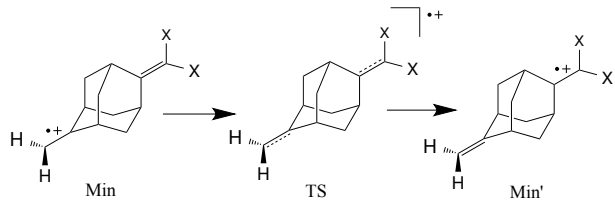}
\caption{Intramolecular electron transport in the \protect\gls{BMA} 
(X = H) and \protect\gls{MIA} (X = CH$_{\text{3}}$) cations \cite{Izmaylov:2011/jcp/234106}.
The Min and Min' structures correspond to the PES minima of the donor and acceptor states, respectively.}
\label{fig:IET}
\end{figure}

To assess performance of \gls{NFGR} for modelling molecular processes using the $H_{\rm QVC}$ Hamiltonian we consider  
intramolecular electron transport in the \gls{BMA} and \gls{MIA} cations (Fig.~\ref{fig:IET}).  
\gls{BMA} and \gls{MIA} contain two unsaturated elements connected by the adamantane cage.  
In \gls{BMA}, the potential minima Min and Min' (Fig.~\ref{fig:IET}) have the same energy and frequencies by symmetry 
while in \gls{MIA} the minimum Min is higher in energy than Min' and vibrational frequencies of the donor and acceptor states 
are different. In both systems modes participating in the intramolecular electron transport can be categorized as either 
tuning modes [$x_{A,i}\ne 0$ in \eq{eq:AcceptRot}] of A$_{\text{1}}$ symmetry or coupling modes 
[$\gamma_i\ne 0$ in \eq{eq:QVC}] of A$_{\text{2}}$ symmetry \cite{Izmaylov:2011/jcp/234106}.  
\gls{BMA} and \gls{MIA} are good candidates for 
the perturbative \gls{NFGR} treatment because the energy scale of the 
interstate coupling, $V_{AD}$ in \eq{eq:Ham}, is small compare to that 
of harmonic frequencies. The rigidity of the adamantane cage makes the harmonic 
approximation very accurate for a majority of vibrational modes in both molecules. 

To obtain adiabatic input for the diabatization we optimized geometry of 
 two ground state minima corresponding to different localizations of 
excessive positive charge as well as the minimum of a \gls{CI} seam using 
the Complete Active Space Self-Consistent Field (CASSCF) method with 
the STO-3G basis and 3 electrons in 4 active space orbitals. These geometric configurations were 
used to evaluate the Hessians in the ground state 
minima and the corresponding Franck-Condon regions of the excited state as well as the gradient difference and 
non-adiabatic coupling vectors at the \gls{CI} seam minimum. 
The resulting diabatic \gls{LVC} and \gls{QVC} Hamiltonians for both molecules are provided in Supplemental Information. 

\begin{figure}
\includegraphics[width=0.5\textwidth]{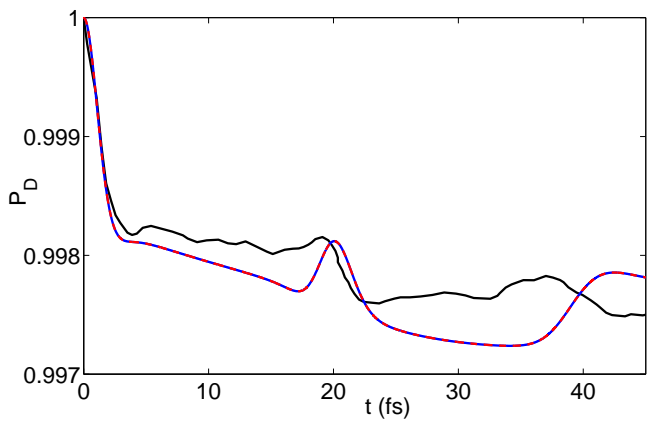}
\caption{Population dynamics for \protect\gls{BMA}: \protect\gls{vMCG} dynamics with 
exact Hamiltonian and 24 Gaussian basis functions (solid black), 
\protect\gls{NFGR} with \protect\gls{QVC} (dashed blue) and \gls{LVC} (dashed red) Hamiltonians.}
\label{fig:BMA}
\end{figure}

Figure \ref{fig:BMA} shows the population dynamics of the \gls{BMA} cation starting from the Boltzmann 
distribution of the uncoupled donor state ($\bff{x_{G}}=0$ and $\mathbf{\Omega}_G=\mathbf{\Omega}_A$) 
using the \gls{NFGR} and \gls{vMCG}\cite{Mendive2013:thesis} approaches.\cite{note:T} 
The perturbative results agree very well with those from the \gls{vMCG} method. 
The \gls{NFGR} approach with both \gls{LVC} and \gls{QVC} Hamiltonians gives the same results for 
\gls{BMA} because all quadratic couplings $\Theta_{i,j}$ are very small and the molecular symmetry does not allow for 
the Duschinsky rotation and frequency difference between the donor and acceptor states. 
The humps in the donor state population around 20 and 40 fs in Fig.~\ref{fig:BMA} 
are due to Rabi like oscillations 
between the ground vibrational state of the donor and the first excited vibrational states of the acceptor 
along the coupling mode with the largest $\gamma_i$. 
The origin of the humps has been confirmed by considering the population dynamics of the modified Hamiltonian where
all but the largest coupling coefficient $\gamma_i$ were zeroed. Population dynamics with the modified Hamiltonian  
had oscillations with the time-scale corresponding to that for the humps in the unmodified Hamiltonian calculation. 
Small discrepancy between \gls{vMCG} and \gls{NFGR} results is attributed to incomplete basis 
convergence in the \gls{vMCG} calculation. 

\begin{figure}
\includegraphics[width=0.5\textwidth]{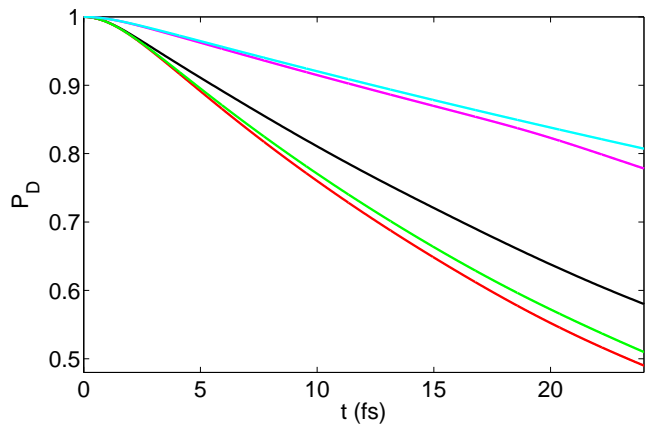}
\caption{Population dynamics for \protect\gls{MIA}: \protect\gls{MCTDH} with the 29D reduced \protect\gls{QVC} Hamiltonian (black), \protect\gls{NFGR} with the 29D reduced \protect\gls{QVC} Hamiltonian (red), \protect\gls{NFGR} with the full \protect\gls{QVC} Hamiltonian (green), \protect\gls{NFGR} with the full \protect\gls{QVC} Hamiltonian without quadratic couplings (light blue), \protect\gls{NFGR} with the full \protect\gls{LVC} Hamiltonian (violet).}
\label{fig:MIA}
\end{figure}

The \gls{MIA} cation is less symmetric than the \gls{BMA} cation 
and its \gls{QVC} Hamiltonian contains significant contributions from the Duschinsky rotation and quadratic couplings.
Due to the greater number of nuclear DOF in the \gls{MIA} cation compared to that in the \gls{BMA} cation
obtaining converged \gls{vMCG} results becomes a difficult task. To assess the 
quality of the \gls{NFGR} population dynamics for \gls{MIA} we compared it with that from the \gls{MCTDH} method
using a reduced 29 dimensional QVC Hamiltonian and the initial  Boltzmann density distribution
in the minimum of the donor state.\cite{note:T} The modes of the reduced 
Hamiltonian were selected so that \gls{NFGR} dynamics of the reduced model is very close to that of the full 96D 
QVC Hamiltonian (see Fig.~\ref{fig:MIA}). 
Figure \ref{fig:MIA} illustrates that NFGR results agree well with those of the MCTDH approach 
and the \gls{QVC} model gives significantly 
different population dynamics than the \gls{LVC} model. The difference in dynamics of the \gls{LVC} and \gls{QVC} 
models has been further separated in two parts arising from the Duschinsky effect 
and the quadratic coupling term. When the Duschinsky effect is included there is a slight decrease 
in the population transfer due to a decrease in \gls{FC} overlaps. 
Adding the quadratic coupling term shows a significant increase in population transfer due to overall increase 
of the coupling between the donor and acceptor states. 


\section{Concluding remarks}
\label{sec:con}
We have developed a perturbative \gls{NFGR} formalism for modelling the electronic population dynamics in large molecules 
parameterized by the \gls{QVC} Hamiltonian. 
The \gls{QVC} model enables different frequencies and Duschinsky 
rotations between the ground, donor and acceptor states.  
We used 2D \gls{QVC} models to explore effects of rotations and shifts in the ground and acceptor states on 
the electronic population dynamics. Analytical expressions obtained in the \gls{NFGR} treatment 
 allowed us to obtain detailed understanding of factors that affect excited state population dynamics. 
 The \gls{NFGR} method has been also applied to the \gls{BMA} and \gls{MIA} cations to illustrate 
 capabilities of the approach and importance of the \gls{QVC} parametrization for the \gls{MIA} cation. 

One of the main advantages of the \gls{NFGR} method is its account for nuclear quantum effects through
the exact quantum treatment of the unperturbed multidimensional harmonic oscillator model. 
Moreover, in contrast to wave-packet approaches, the \gls{NFGR} method implicitly 
operates in a complete nuclear basis because the perturbative formalism enables the exact summation over  
the complete set of the nuclear states. This feature gives us exact short term dynamics and makes \gls{NFGR} 
a complementary approach to wave-packet techniques where convergence with respect to the nuclear basis is 
a common concern. Another useful feature of \gls{NFGR} is its computational efficiency, \gls{NFGR} can 
easily treat thousands of nuclear \gls{DOF}, 
which will allow us to explore electronic transitions in large molecules and materials.  
With only the results of electronic structure calculations the \gls{NFGR} approach can provide a quick estimate of 
the electronic transition dynamics without simulating computationally expensive wave-packet quantum dynamics. 

Two main limitations of the \gls{NFGR} approach 
are use of the perturbative approximation and fixed parametrized \gls{QVC} Hamiltonian 
instead of a nuclear Hamiltonian generated on-the-fly.\cite{Saita:2012di,BenNun:2002tx} 
The perturbative approximation provides accurate dynamics either for weak couplings or for short times. Its range 
of applicability can be somewhat increased by putting developed \gls{NFGR} correlation 
functions within the generalized master equation framework.~\cite{Book/Breuer:2002}  
Even better approach to both \gls{NFGR} limitations is in a framework of a recently proposed perturbative 
spawning method\cite{Izmaylov:2013/jcp/104115} which can potentially use model Hamiltonians parametrized on-the-fly
at every time step.  The perturbative spawning approach consists of using the 
\gls{vMCG}\cite{Worth:2004/fd/307} 
method to propagate the nuclear wave function and a perturbative approach to determine whether the number of basis
functions is enough to maintain an adequate level of accuracy. The \gls{NFGR} development for the \gls{QVC} 
Hamiltonian can be incorporated in the pertubative spawning method to estimate the difference in electronic 
population transfer using a complete nuclear basis and a finite basis involved in the actual propagation. 
If the difference is greater than a given threshold then new 
wave-packets are spawned. The structure of the spawning procedure is similar to that proposed by Martinez {\it et al.}
in the {\it ab initio} multiple spawning method\cite{BenNun:2002tx} 
but the spawning criterion in the perturbative spawning approach 
rigorously follows from perturbation theory estimate. 
The perturbative spawning with the \gls{NFGR} treatment of the \gls{QVC} 
Hamiltonians will combine best qualities of both variational and pertubative 
approaches and its implementation will be a subject of future work.
 
\section{Acknowledgments}
We thank David Mendive-Tapia for providing numerical data for the on-the-fly \gls{vMCG} simulation of \gls{BMA}. 
AFI acknowledges funding from the Natural Sciences and
Engineering Research Council of Canada (NSERC) through the Discovery Grants
Program. LJD thanks the European Union Seventh Framework Programme
(FP7/2007-2013) for financial support under grant agreement PIOF-GA-2012-332233.

\section{Appendix}

To evaluate the trace in Eq.~(\ref{eq:fTr}) we use a method developed by Kubo and Toyozawa\cite{Kubo:1955/ptp/160} 
which involves writing the exponential operators of Eq.~(\ref{eq:fTr}) using position eigen-functions 
\bea
f(t',t'')&=&\int\limits_{-\infty}^{\infty} \mathrm{d}\bff{q}\int\limits_{-\infty}^{\infty} \mathrm{d}\bff{q}'\int\limits_{-\infty}^{\infty} \mathrm{d}\bff{q}''\int\limits_{-\infty}^{\infty} \mathrm{d}\bff{q}'''\bra{\bff{q}}e^{-\beta H_{G}}\ket{\bff{q}'}\nn\\
&&\times\bra{\bff{q}'}e^{iH_{D}t'}\ket{\bff{q}''} V_{DA} \bra{\bff{q}''}e^{-iH_{A}(t'-t'')}\ket{\bff{q}'''}V_{AD}\nn\\
&&\times \bra{\bff{q}'''}e^{-iH_{D}t''}\ket{\bff{q}}\left(\int\limits_{-\infty}^{\infty} \mathrm{d}\bff{q} \bra{\bff{q}}e^{-\beta H_{G}}\ket{\bff{q}}\right)^{-1},\label{eq:int}
\eea
and then integrating over the nuclear \gls{DOF}. 
Matrix elements of all exponential operators can be evaluated analytically, because
 for any quadratic Hamiltonian
\be
H_Q = \sum_{i=1}^{N}\frac{1}{2}p_i^{2}+\frac{1}{2}\Omega_{i}^{2}(q_{i}-x_i)^{2},
\ee
we can write\cite{Kubo:1955/ptp/160}
\begin{equation}\label{eq:nDHO}
\begin{aligned}
\frac{\left<\bff{q}\left|e^{-\beta H_Q}\right|\bar{\bff{q}}\right>}{{\rm Tr}[e^{-\beta H_Q}]} &=\left\{\det\left[2\pi\frac{\sinh(\beta \bs{\Omega})}{\bs{\Omega}}\right]\right\}^{-1/2}\exp\Bigg\{-\frac{1}{4}\\
&\times\Bigg[(\bff{q}+\bar{\bff{q}}-2\bff{x})\bs{\Omega}\tanh \left(\frac{\beta \bs{\Omega}}{2}\right)(\bff{q}+\bar{\bff{q}}-2\bff{x})\\
&+(\bff{q}-\bar{\bff{q}})\bs{\Omega}\coth\left(\frac{\beta \bs{\Omega}}{2}\right) (\bff{q}-\bar{\bff{q}})\Bigg]\Bigg\}\\
\end{aligned}
\end {equation}
for real $\beta$ and then use analytic continuation of \eq{eq:nDHO} for the matrix element of the time 
propagator $\left<\bff{q}\left|e^{-it H_Q}\right|\bar{\bff{q}}\right>$. 
In \eq{eq:nDHO} $\bff{q},\bff{\bar{q}}$, and $\bff{x}$ are $N$-dimensional vectors, and 
$\bs\Omega$ is an $N$-dimensional frequency matrix.
Using this consideration to express the matrix elements of the exponential operators in Eq.~(\ref{eq:int}) 
we write the time-correlation function as
\begin{widetext}
\bea \label{eq:Kubo}
f(t',t'') &=& \det[\bff{S_{G}S_{D}}(t')\bff{S_{A}S_{D}}^{*}(t'')]^{-1/2}\int_{-\infty}^{\infty} \mathrm{d}\bff{q}\int_{-\infty}^{\infty} \mathrm{d}\bff{q}'\int_{-\infty}^{\infty} \mathrm{d}\bff{q}''\int_{-\infty}^{\infty} \mathrm{d}\bff{q}''' e^{-i\Delta E(t'-t'')} \notag\\ 
&&\times\exp\Big\{-\frac{1}{4}\Big[(\bff{q}+\bff{q}'-2\bff{x}_{G})^{T}\bff{T_{G}}(\bff{q}+\bff{q}'-2\bff{x}_{G})+(\bff{q}-\bff{q}')^{T}\bff{C_{G}}(\bff{q}-\bff{q}') \notag\\
&&\hspace{1cm}+(\bff{q}'+\bff{q}'')^{T}\bff{T_{D}}(t')(\bff{q}'+\bff{q}'')+(\bff{q}'-\bff{q}'')^{T}\bff{C_{D}}(t')(\bff{q}'-\bff{q}'')\notag\\ 
&&\hspace{1cm}+(\bff{q}''+\bff{q}'''-2\bff{x}_{A})^{T}\bff{T_{A}}(\bff{q}''+\bff{q}'''-2\bff{x}_{A})+(\bff{q}''-\bff{q}''')^{T}\bff{C_{A}}(\bff{q}''-\bff{q}''')\notag\\
&&\hspace{1cm}+(\bff{q}'''+\bff{q})^{T}\bff{T_{D}}^{*}(t'')(\bff{q}'''+\bff{q})+(\bff{q}'''-\bff{q})^{T}\bff{C_{D}}^{*}(t'')(\bff{q}'''-\bff{q})\Big]\Big\}\nn\\
&&\times V(\bff{q}'')V(\bff{q}''')\left\{\det(\bff{S_{G}})^{-1/2} \int_{-\infty}^{\infty}\mathrm{d}\bff{q}e^{-(\bff{q}-\bff{x_{G}})\bff{T_{G}}(\bff{q}-\bff{x_{G}})}\right\}^{-1},
\eea
where
\bea
\bff{T_{G}}\phantom{(t')}  &=&\bff{J^{(G)}}^{T}\bs\Omega_{G}\tanh \left(\frac{\beta \bs\Omega_{G}}{2}\right)\bff{J^{(G)}}, \\
\bff{C_{G}}\phantom{(t')}  &=& \bff{J^{(G)}}^{T}\bs\Omega_{G}\coth\left(\frac{\beta \bs\Omega_{G}}{2}\right)\bff{J^{(G)}}, \\
\bff{S_{G}}\phantom{(t')}  &=& 2\pi\frac{\sinh(\beta \bs\Omega_{G})}{\bs\Omega_{G}}, \\
\bff{T_{D}}(t')  &=&\bs\Omega_{D}\tanh \left(\frac{-it' \bs\Omega_{D}}{2}\right),\\
\bff{C_{D}}(t')  &=& \bs\Omega_{D}\coth\left(\frac{-it' \bs\Omega_{D}}{2}\right), \\
\bff{S_{D}}(t')  &=& 2\pi\frac{\sinh(-it' \bs\Omega_{D})}{\bs\Omega_{D}}, \\
\bff{T_{A}}\phantom{(t')}&=&\bff{J^{(A)}}^{T}\bs\Omega_{A}\tanh \left[\frac{i(t'-t'') \bs\Omega_{A}}{2}\right]\bff{J^{(A)}},\\
\bff{C_{A}} &=& \bff{J^{(A)}}^{T}\bs\Omega_{A}\coth\left[\frac{i(t'-t'') \bs\Omega_{A}}{2}\right]\bff{J^{(A)}}, \\
\bff{S_{A}}&=&2\pi \frac{\sinh[i(t'-t'') \bs\Omega_{A}]}{\bs\Omega_{A}}, \\ 
V(\bff{q}) &=& \bff{q}^{T}\bs\Theta \bff{q}+\bs\gamma^{T}\bff{q}+\Delta_{DA} ,
\eea
and all nuclear coordinates correspond to the donor state normal modes.  Using Gaussian 
integration in \eq{eq:Kubo} we obtain the analytic expression for the $f(t',t'')$ function
\bea
f(t',t'')&=&\phi(t',t'')\chi(t',t'')\lambda(t',t'') e^{-i\Delta E(t'-t'')}\label{eq:f}\\
\phi(t',t'')&=&\det\Big\{\left[(2\pi)^2\sinh(\beta \bs\Omega_{D}/2)\right]^{-2}\bff{S_{G}S_{D}(t')S_{A}S_{D}^*(t'')A_1A_{2}A_{3}A_{4}}\Big\}^{-1/2},\label{eq:phi}\\
\chi(t',t'')&=&\exp\left[\frac{1}{4}\bff{b_{4}}^{T}\bff{A_{4}}^{-1}\bff{b_{4}}+c_{4}\right],\label{eq:chi}\\
\lambda(t',t'')&=&\left[\fo{2}\Tr(\bff{A_{4}}^{-1}\bs\Theta_{4})+\bff{k_{1}}^{T}\bs\Theta_{4}\bff{k_{1}}-\bs\gamma_{4}^{T}\bff{k_{1}}\right]\Delta_{DA}-\left[\bff{k_{1}}^{T}\bs\Theta_{4}\bff{k_{1}}\bs\gamma ^{T}\bff{k_{1}}+\bff{k_{1}}^{T}\bs\Theta \bff{k_{1}}\bs\gamma_{4}^{T}\bff{k_{1}}\right]\nn \\
&&+\left[\fo{2}\Tr(\bff{A_{4}}^{-1}\bs\Theta)+\bff{k_{1}}^{T}\bs\Theta \bff{k_{1}}-\bs\gamma ^{T}\bff{k_{1}}\right]\Delta_{DA}'-\left[\bs\gamma ^{T}\bff{A_{4}}^{-1}\bs\Theta_{4}\bff{k_{1}}+\bs\gamma_{4}^{T}\bff{A_{4}}^{-1}\bs\Theta \bff{k_{1}}\right]\nn\\
&&+\fo{2}\left[\Tr(\bff{A_{4}}^{-1}\bs\Theta )\bff{k_{1}}^{T}\bs\Theta_{4}\bff{k_{1}}+\Tr(\bff{A_{4}}^{-1}\bs\Theta_{4})\bff{k_{1}}^{T}\bs\Theta \bff{k_{1}}\right]+\bff{k_{1}}^{T}\bs\Theta \bff{k_{1}}\bff{k_{1}}^{T}\bs\Theta_{4}\bff{k_{1}}+\Delta_{DA}\Delta_{DA}'\nn\\
&&-\fo{2}\left[\Tr(\bff{A_{4}}^{-1}\bs\Theta_{4})\bs\gamma ^{T}\bff{k_{1}}+\Tr(\bff{A_{4}}^{-1}\bs\Theta)\bs\gamma_{4}^{T}\bff{k_{1}}\right]+2\bff{k_{1}}^{T}\bs\Theta_{4}\bff{A_{4}}^{-1}\bs\Theta \bff{k_{1}}+\bff{k_{1}}^{T}\bs\gamma\bs\gamma_4^{T}\bff{k_{1}}\nn\\
&&+\fo{2}\Tr(\bff{A_{4}}^{-1}\bs\Theta \bff{A_{4}}^{-1}\bs\Theta_{4})+\fo{4}\Tr(\bff{A_{4}}^{-1}\bs\Theta )\Tr(\bff{A_{4}}^{-1}\bs\Theta_{4})+\fo{2}\bs\gamma_{4}^{T}\bff{A_{4}}^{-1}\bs\gamma,\label{eq:lambda}
\eea
\bea
\bff{A_1}&=&\bff{T_{G}}+\bff{C_{G}}+\bff{T_{D}(t'')}+\bff{C_{D}(t'')}, \\
\bff{A_{2}}&=&\frac{1}{4}\left[-(\bff{T_{G}}-\bff{C_{G}})\bff{A_1}^{-1}(\bff{T_{G}}-\bff{C_{G}})+\bff{T_{D}(t')}+\bff{C_{D}(t')}+\bff{T_{G}}+\bff{C_{G}}\right],\\
\bff{A_{3}}&=&\frac{1}{4}\Big\{-\frac{1}{4}\left[\bff{T_{D}(t')}-\bff{C_{D}(t')}\right]\bff{A_{2}}^{-1}\left[\bff{T_{D}(t')}-\bff{C_{D}(t')}\right]+\bff{T_{D}(t')}+\bff{C_{D}(t')}\nn\\
&&+\bff{T_{A}}+\bff{C_{A}}\Big\},\\
\bff{A_{4}}&=&\frac{1}{4}\Big\{-\frac{1}{4}[\bff{T_{D}(t'')}-\bff{C_{D}(t'')}]\bff{A_1}^{-1}[\bff{T_{D}(t'')}-\bff{C_{D}(t'')}]\big[\bff{A_1}^{-1}(\bff{T_{G}}-\bff{C_{G}})\bff{A_{2}}^{-1}\times\nn\\
&&\hspace{1cm}(\bff{T_{G}}-\bff{C_{G}})+4\big]-\bff{K}\bff{A_{3}}^{-1}\bff{K}+\bff{T_{D}(t'')}+\bff{C_{D}(t'')}+\bff{T_{A}}+\bff{C_{A}}\Big\},
\eea
\bea
%
\bff{K}&=&-\frac{1}{8}[\bff{T_{D}(t'')}-\bff{C_{D}(t'')}]\bff{A_1}^{-1}(\bff{T_{G}}-\bff{C_{G}})\bff{A_{2}}^{-1}
[\bff{T_{D}(t')}-\bff{C_{D}(t')}] \nn\\
&&-\frac{1}{2}\bff{T_{A}}+\frac{1}{2}\bff{C_{A}},\\
\bff{b_{4}}&=&\frac{1}{4}[\bff{T_{D}(t'')}-\bff{C_{D}(t'')}]\Big\{\bff{A_{2}}^{-1}[\bff{T_{G}}-\bff{C_{G}}][\bff{1}-\bff{A_1}^{-1}(\bff{T_{G}}-\bff{C_{G}})]-4\Big\}\times\nn\\
&&\bff{A_1}^{-1}\bff{T_{G}}\bff{x_{G}}+\frac{1}{2}\bff{K}\bff{A_{3}}^{-1}\bff{k_{2}}+\bff{T_{A}}\bff{x_{A}},\\
c_{4}&=&\frac{1}{4}\bff{k_{2}}^{T}\bff{A_{3}}^{-1}\bff{k_{2}}+\frac{1}{4}\bff{k_3}^{T}\bff{A_{2}}^{-1}\bff{k_3}+\bff{x_{G}}^{T}\left[\bff{T_{G}}\bff{A_1}^{-1}\bff{T_{G}}-\bff{T_{G}}\right]\bff{x_{G}}-\bff{x_{A}}^{T}\bff{T_{A}}\bff{x_{A}},\\
\bs\Theta_{4}&=&\frac{1}{4}\bff{K}\bff{A_{3}}^{-1}\bs\Theta \bff{A_{3}}^{-1}\bff{K}, \\
\bs\gamma_{4}&=&\frac{1}{2}\bff{K}\bff{A_{3}}^{-1}\left[\bs\Theta \bff{A_{3}}^{-1}\bff{k_{2}}+\bs\gamma\right],\\
\Delta_{DA}'&=&\frac{1}{2}\Tr(\bs\Theta \bff{A_{3}}^{-1})+\Delta_{DA} +\frac{1}{4}\bff{k_{2}}^{T}\bff{A_{3}}^{-1}\left[\bs\Theta \bff{A_{3}}^{-1}\bff{k_{2}}+2\bs\gamma\right],\\
\bff{k_{1}}&=&-\frac{1}{2}\bff{A_{4}}^{-1}\bff{b_{4}},\\
\bff{k_{2}}&=&-\frac{1}{4}[\bff{T_{D}(t')}-\bff{C_{D}(t')}]\bff{A_{2}}^{-1}\bff{k_3}+\bff{T_{A}}\bff{x_{A}},\\
\bff{k_3}&=&\left[\bff{1}-(\bff{T_{G}}-\bff{C_{G}})\bff{A_1}^{-1}\right]\bff{T_{G}}\bff{x_{G}}.
\eea
\end{widetext}
 The function $f(t',t'')$ contains all parameters of the \gls{QVC} and ground state Hamiltonians as 
 well as temperature of the initial Boltzmann distribution. 
 Although the equations for the $f(t',t'')$ components $\phi,\chi,$ and $\lambda$ 
 are lengthy their physical meaning is quite transparent. 
 The $\lambda$ term contains the quadratic dependence on the inter electronic couplings, as  
 expected considering the second order perturbation theory used in deriving $f(t',t'')$.
 Time-dependence associated with the linear and quadratic coupling terms in $\lambda$ accounts 
 for the energy difference between levels coupled by these terms.
 If the linear and quadratic couplings are set to zero the $\lambda$ term reduces to $\Delta_{DA}^2$ 
 as in the simpler Fermi golden rule expression of Borrelli {\it et al.}\cite{Borrelli:2011/PCCP/4420} 
 Another contribution to the overall coupling of the donor and acceptor states is the Franck-Condon 
 overlap between nuclear states of corresponding harmonic wells. This contribution is modelled
 by the $\chi$ term that depends on spatial shifts between minima of electronic states and orientation 
of normal modes corresponding to different electronic states.
To fully account for all contributions into energy differences between coupled harmonic nuclear states of 
different diabatic states there are two additional terms in the $f(t',t'')$ expression: 
First, the $\exp[-i\Delta E(t'-t'')]$ term introduces energy difference between minima of the donor and acceptor 
states. Second, the $\phi$ term accounts for the difference between vibrational frequencies 
of the donor and acceptor states. If $\bs\Omega_{D}=\bs\Omega_{A}$ then $\phi=1$ as in  
the \gls{NFGR} formalism for the \gls{LVC} Hamiltonian where there is
 only one set of harmonic frequencies for all electronic states. 

\begin{thebibliography}{41}
\bibitem [{not()}]{note:T}%
  \BibitemOpen
  \href@noop {} {}\bibinfo {note} {We have used very small but finite values of
  temperature ($\sim 10^{-5}$ a.u.) in the Boltzmann distribution for NFGR
  simulations to make them comparable with variational calculations at zero
  temperature.}\BibitemShut {Stop}%
\end{thebibliography}
%

 \end{document}